\documentclass[fleqn,usenatbib]{mnras}

\usepackage{newtxtext,newtxmath}
\usepackage[T1]{fontenc}
\usepackage[utf8]{inputenc}

\DeclareRobustCommand{\VAN}[3]{#2}
\let\VANthebibliography\thebibliography
\def\thebibliography{\DeclareRobustCommand{\VAN}[3]{##3}\VANthebibliography}

\usepackage{graphicx}

\newcommand{\ud}{\mathrm{d}}

\title[Simulations of Oxygen Shell Burning]{3D Simulations of Oxygen Shell Burning with and without Magnetic Fields}

\author[Varma \& M\"uller]{
Vishnu Varma$^{1}$\thanks{E-mail: 
vishnu.rvejayan@monash.edu}
and
Bernhard M\"uller$^{1,2}$\thanks{E-mail: 
bernhard.mueller@monash.edu}
\\
$^{1}$
School of Physics and Astronomy, 10 College Walk, Monash University, Clayton, VIC 3800, Australia\\
$^{2}$
ARC Centre of Excellence for Gravitational Wave Discovery -- OzGrav
}

\date{Accepted XXX. Received YYY; in original form ZZZ}

\pubyear{2020}

\begin{document}
\label{firstpage}
\pagerange{\pageref{firstpage}--\pageref{lastpage}}
\maketitle

\begin{abstract}
We present a first 3D magnetohydrodynamic (MHD) simulation of
convective oxygen and neon shell burning in a non-rotating $18\,
M_\odot$ star shortly before core collapse to study the generation of
magnetic fields in supernova progenitors. We also run a purely
hydrodynamic control simulation to gauge the impact of the magnetic
fields on the convective flow and on convective boundary mixing. After
about 17 convective turnover times, the magnetic field is approaching
saturation levels in the oxygen shell with an average field strength
of $\mathord{\sim}10^{10}\, \mathrm{G}$, and does not reach kinetic
equipartition. The field remains dominated by small to medium scales,
and the dipole field strength at the base of the oxygen shell is only
$10^{9}\, \mathrm{G}$.  The angle-averaged diagonal components of the
Maxwell stress tensor mirror those of the Reynolds stress tensor, but
are about one order of magnitude smaller. The shear flow at the
oxygen-neon shell interface creates relatively strong fields parallel
to the convective boundary, which noticeably inhibit the turbulent
entrainment of neon into the oxygen shell. The reduced ingestion of
neon lowers the nuclear energy generation rate in the oxygen shell and
thereby slightly slows down the convective flow. Aside from this
indirect effect, we find that magnetic fields do not appreciably alter
the flow inside the oxygen shell. We discuss the implications of our
results for the subsequent core-collapse supernova and stress the need
for longer simulations, resolution studies, and an investigation of
non-ideal effects for a better understanding of magnetic fields in
supernova progenitors.
\end{abstract}

\begin{keywords}
stars: massive -- stars: magnetic fields -- stars:interiors
-- MHD -- convection -- turbulence
\end{keywords}

\section{Introduction}
\label{sec:intro} 

Rigorous three-dimensional (3D) simulations of neutrino-driven core-collapse supernovae have become highly successful in recent years \citep[e.g.,][]{melson_15b,lentz_15,takiwaki_14,burrows_19}, and have made clear headway in explaining the properties of supernova explosions and the compact objects born in these events \citep{mueller_17,Mueller2019,Burrows2020b,Powell2020, Bollig2020, mueller_20}. 
The latest 3D models are able to reproduce a range of explosion energies up to $10^{51}\mathrm{erg}$ \citep{Bollig2020}, and yield neutron star birth masses, kicks, and spins largely compatible with the population of observed young pulsars \citep{Mueller2019,Burrows2020b}. 

A number of ingredients have contributed, or have
the potential to contribute,  to make modern 
neutrino-driven explosion models more robust.
Various microphysical effects such as
reduced neutrino scattering opacities due
to nucleon strangeness \citep{melson_15b} and nucleon correlations at high densities \citep{horowitz_17,Bollig2017,burrows_18}, muonisation \citep{Bollig2017}, and large effective nucleon masses \citep{yasin_20} can be conducive to neutrino-driven shock revival. 

In addition, a particularly important turning point has been the advent of 3D \emph{progenitor} models and the recognition that
asphericities seeded prior to collapse can precipitate ``perturbation-aided'' neutrino-driven explosions
\citep{couch_13,couch_15, mueller_15a, mueller_16b,mueller_17,mueller_20,Bollig2020}. In perturbation-aided explosions, the moderately subsonic solenoidal velocity perturbation
in active convective shells at the pre-collapse stage
with Mach numbers of order $\mathord{\sim}0.1$
\citep{Collins2018} are transformed into strong
density and pressure perturbations at the
shock \citep{mueller_15a,takahashi_14,takahashi_16,abdikamalov_20}, and effectively strengthen the violent non-radial flow behind the shock
that develops due to convective
instability \citep{herant_94,burrows_95,janka_95,janka_96} and the standing accretion shock instability \citep{blondin_03,foglizzo_07}, thereby supporting neutrino-driven shock revival.

The discovery of perturbation-aided explosions has greatly enhanced interest in multi-dimensional models of late convective burning stages in massive stars. Several studies have by now followed the immediate pre-collapse phase of silicon and/or oxygen shell burning to collapse in 3D
\citep{couch_15,mueller_16b,mueller_16c,Mueller2019,Yoshida2019,Yadav2020,McNeill2020,yoshida_20}. In addition, there has been a long-standing strand of research since the 1990s \citep[e.g.,][]{bazan_94,bazan_97,meakin_06,meakin_07,meakin_07b,mueller_16c,jones_17,cristini_17,cristini_19} into the detailed behaviour of stellar convection during advanced burning stages with a view to the \emph{secular}  impact of 3D effects not captured by spherically symmetric stellar evolution models based on mixing-length theory \citep{Biermann1932,Boehm1958}. The details of convective boundary mixing by processes such as quasi-steady turbulent entrainment \citep{Fernando1991,strang_01,meakin_07}
or violent shell mergers \citep{mocak_18,Yadav2020} have received particular attention for their potential to alter the core structure of massive stars and hence affect the dynamics and final nucleosynthesis yields of the subsequent supernova explosion.

Three-dimensional simulations of convection during advanced burning stages have so far largely disregarded two important aspects of real stars -- rotation and magnetic fields. The effects of rotation have been
touched upon by the seminal work of \citet{kuhlen_03}, but more recent studies \citep{arnett_10,chatzopoulos_16} have been limited to axisymmetry (2D). Magnetohydrodynamic (MHD) simulations of convection, while common and mature in the context of the Sun \citep[for a review see, e.g.,][]{charbonneau_14}  have yet to be performed for advanced burning stages of massive stars.

Simulations of magnetoconvection during the late burning stages, both in rotating and non-rotating stars, are a big desideratum for several reasons.
Even in slowly rotating massive stars, magnetic fields may have a non-negligible impact on the dynamics of neutrino-driven explosions \citep{obergaulinger_14,Mueller2020}, and although efficient field amplification processes operate
in the supernova core \citep{endeve_12,Mueller2020},
it stands to reason that memory of the initial fields may not be lost entirely, especially for strong fields in the progenitor and early explosions. A better understanding of the interplay between convection, rotation, and magnetic fields in supernova progenitors
is even more critical for the magnetorotational explosion scenario \citep[e.g.,][]{burrows_07a,Winteler2012a,Mosta2014,moesta_18,Obergaulinger2020,Obergaulinger2020b,kuroda_20,aloy_21}, which probably 
explains rare, unusually energetic ``hypernovae'' with energies of up to $\mathord{\sim}10^{52}\,\mathrm{erg}$.
Again, even though the requisite strong magnetic
fields may be generated after collapse by amplification processes like the magnetorotational insstability \citep{balbus_91,akiyama_03} or an $\alpha$-$\Omega$
dynamo in the proto-neutron star \citep{duncan_92,thompson_93,Raynaud2020}, a robust understanding of magnetic
fields during late burning is indispensable for reliable
hypernova models on several heads. For sufficiently
strong seed fields in the progenitor, the initial
field strengths and geometry could have a significant
impact on the development of magnetorotational explosions after collapse \citep{bugli_20,aloy_21}. Furthermore, our understanding of evolutionary pathways towards hypernova explosions
\citep{woosley_06,yoon_05,yoon_10,cantiell_07,aguilera_18,aguilera_20}
is intimately connected 
with the effects of magnetic fields
on angular momentum transport in stellar interiors
\citep{spruit_02,heger_05,fuller_19,takahashi_20}.

Beyond the impact of magnetic fields on the pre-supernova
evolution and the supernova itself, the interplay of convection, rotation, and magnetic fields is obviously relevant to the origin of neutron star magnetic fields as well. It still remains to be explained what shapes the distribution of magnetic fields among young pulsars, and why some neutron stars are born as magnetars with extraordinarily strong dipole fields of up to $10^{15}\,\mathrm{G}$
\citep{olausen_14,tauris_15,Kaspi2017,Enoto2019}. Are these strong fields of fossil origin \citep{ferrario_05,Ferrario2009,schneider_20} or generated by
dynamo action during or after the supernova \citep{duncan_92,thompson_93}?
Naturally, 3D MHD simulations of  the late burning stages cannot comprehensively answer all of these questions. In order to connect to
observable magnetic fields of neutron stars, an integrated approach is required that combines stellar evolution over secular time scales, 3D stellar hydrodynamics, supernova modelling, and local simulations, and also addresses aspects like field burial \citep{vigano_13,torres_16} and the long-time evolution
of magnetic fields \citep{aguilera_08,de_grandis_20}. 
Significant obstacles need to be overcome until  one
can construct a pipeline
from 3D progenitor models to 3D supernova simulations and beyond.
However, 3D MHD simulations of convective burning can already address meaningful questions despite the complexity of the overall problem.
As in the purely hydrodynamic case, the first step must be to understand the principles governing magnetohydrodynamic convection
during the late burning stages based on somewhat
idealised simulations that are broadly representative of 
the conditions in convective cores and shells of massive stars.

In this study, we present a first simulation of magnetoconvection
during the final phase of oxygen shell burning using the
ideal MHD approximation. In the tradition of
semi-idealised models of late-stage convection, we use a setup
that falls within the typical range of conditions in the
interiors of massive stars in terms of convective Mach number
and shell geometry \citep[for an overview see][]{Collins2018}
and is not designed as a fully self-consistent
model of any one particular star.
This simulation constitutes a first
step beyond effective 1D prescriptions in stellar evolution models
to predict the magnetic field strength and geometry
encountered in the inner shells of massive stars at the pre-supernova
stage. We also compare to a corresponding non-magnetic model of
oxygen shell convection to gauge the feedback of magnetic fields
on the convective flow with a particular view to two important issues.
First, the efficiency of the ``perturbation-aided'' neutrino-driven mechanism
depends critically on the magnitude of the convective velocities
during shell burning, and  it is important to determine whether
magnetic fields can significantly slow down convective motions
as suggested by some recent simulations of solar convection
\citep{hotta_15}. 
Second, magnetic fields could quantitatively
or qualitatively affect shell growth by turbulent entrainment,
which has been consistently seen in all recent 3D hydrodynamics
simulations of late-stage convection in massive stars.

Our paper is structured as follows. In Section \ref{sec:methods}, we describe the
 numerical methods, progenitor model, and initial conditions
 used in our study and discuss the potential role of non-ideal effects . The results of the simulations are  presented in Section~\ref{sec:results}. We first focus on the strength and geometry of the emerging magnetic field 
and then analyse the impact of magnetic fields on the flow, and in particular
on entrainment at shell boundaries. We summarize our results and discuss their implications in Section~\ref{sec:conclusion}. 

\begin{figure}
    \centering
    \includegraphics[width=\linewidth]{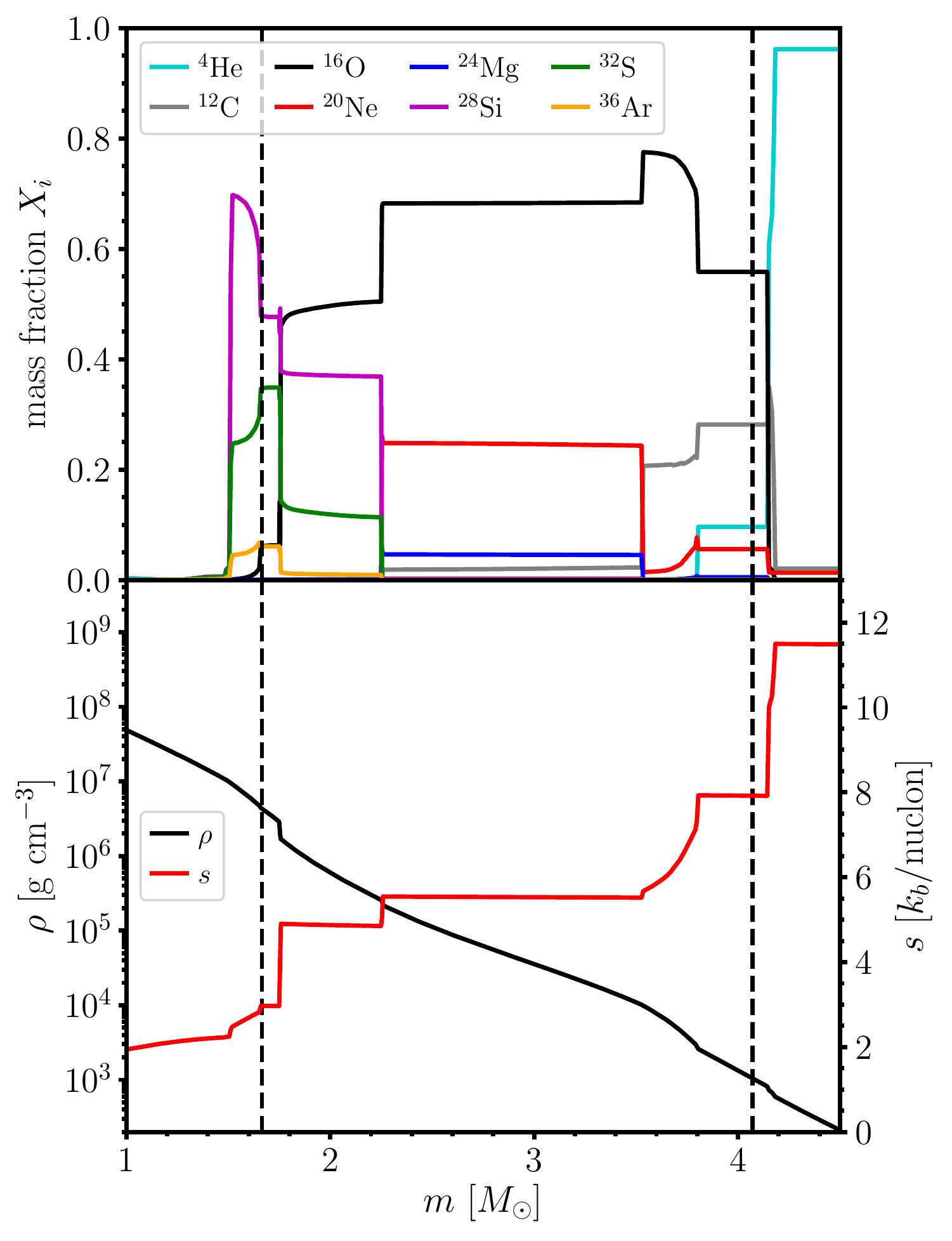}
    \caption{Profiles of selected mass fractions $X_i$ (top), density $\rho$, and specific
    entropy $s$ in the $18\, M_\odot$ \textsc{Kepler} stellar evolution model at the time
    of mapping to \textsc{CoCoNuT}. The boundaries of the simulated
    domain are indicated by dashed vertical lines. Note
    subtle differences to Figure~1 in \citet{mueller_16c}, which
    shows the same stellar evolution model at the onset of collapse.}
    \label{fig:1d_model}
\end{figure}

\section{Numerical Methods and Simulation Setup}
\label{sec:methods} 

We simulate oxygen and neon shell burning 
with and without magnetic fields
in a non-rotating $18\,M_\odot$ solar-metallicity star calculated using the  stellar evolution code \textsc{Kepler} \citep{Weaver1978, Woosley2002, Heger2010}. The same progenitor model has previously been used in the shell convection simulation of \citet{mueller_16c}. 
The structure of the stellar evolution model at
the time of mapping to 3D is illustrated
in Figure~\ref{fig:1d_model}. The model contains
two active convective shells with sufficiently
short turnover times to make time-explicit simulations feasible.
The oxygen shell extends from 
$1.75\, M_\odot$ to $2.25\, M_\odot$ in enclosed
mass and from $3,400\, \mathrm{km}$
to $7,900\,\mathrm{km}$ in radius, immediately
followed further out by the neon shell
out to $3.53\, M_\odot$ in mass and
$27,000\, \mathrm{km}$ in radius.

For our 3D simulations we employ the Newtonian magnetohydrodynamic (MHD) version of the  \textsc{CoCoNuT} code as described in \citet{Mueller2020}.  
The MHD equations are solved in spherical
polar coordinates using the HLLC (Harten-Lax-van Leer-Contact) Riemann solver \citep{Gurski2004a, Miyoshi2005a}. The divergence-free condition $\nabla\cdot\mathbf{B} = 0$ is maintained using a 
modification
of the original hyperbolic divergence cleaning scheme of \citet{Dedner2002} that allows for a variable cleaning speed while still maintaining total energy conservation using an idea similar to \citet{Tricco2016}. Compared to the original cleaning method, we rescale
the Lagrange multiplier $\psi$ to
$\hat{\psi}=\psi/c_\mathrm{h}$, where the cleaning speed $c_\mathrm{h}$ is chosen to be the fast magnetosonic speed. Details of this approach and differences
to \citet{Tricco2016} are discussed in Appendix~\ref{app:eglm}.
The extended system of MHD equations for the density $\rho$, velocity $\mathbf{v}$, magnetic field $\mathbf{B}$, the total energy density $\hat{e}$, mass
fractions $X_i$, and the rescaled Lagrange multiplier $\hat{\psi}$ reads,
\begin{eqnarray}
\partial_t \rho
+\nabla \cdot \rho \mathbf{v}
&=&
0,
\\
\partial_t (\rho \mathbf v)
+\nabla \cdot \left(\rho \mathbf{v}\mathbf{v}-
\frac{\mathbf{B} \mathbf{B}}{4\pi}
+P_\mathrm{t}\mathcal{I}
\right)
&=&
\rho \mathbf{g}
-
\frac{(\nabla \cdot\mathbf{B}) \mathbf{B}}{4\pi}
,
\\
\partial_t {\hat{e}}+
\nabla \cdot 
\left[(e+P_\mathrm{t})\mathbf{u}
-\frac{\mathbf{B} (\mathbf{v}\cdot\mathbf{B})
{-c_\mathrm{h} \hat{\psi} \mathbf{B}}}{4\pi}
\right]
&=&
\rho \mathbf{g}\cdot \mathbf{v}
+
\rho \dot{\epsilon}_\mathrm{nuc}
,
\\
\partial_t \mathbf{B} +\nabla \cdot (\mathbf{v}\mathbf{B}-\mathbf{B}\mathbf{v})
+\nabla  (c_\mathrm{h} \hat{\psi})
&=&0,
\\
\partial_t \hat{\psi}
+c_\mathrm{h} \nabla \cdot \mathbf{B}
&=&-\hat{\psi}/\tau,
\\
\partial_t (\rho X_i)
+\nabla \cdot (\rho  X_i \mathbf{v})
&=&
\rho \dot{X}_i
,
\end{eqnarray}
where $\mathbf{g}$ is the gravitational acceleration, $P_\mathrm{t}$ is the total pressure, $\mathcal{I}$ is the Kronecker tensor, $c_\mathrm{h}$ is the hyperbolic cleaning speed, $\tau$ is the damping time scale for divergence cleaning, and $\dot{\epsilon}_\mathrm{nuc}$ and $\dot{X}_i$ are energy and mass fraction
source terms from nuclear reactions. This
system conserves the volume integral of a
modified total energy density $\hat{e}$,
which also contains the cleaning field $\hat{\psi}$,
\begin{equation}
{\hat{e}}
=\rho \left(\epsilon+\frac{v^2}{2}\right)+\frac{B^2+\hat{\psi}^2}{8\pi},
\end{equation}
where $\epsilon$ is the mass-specific internal energy. 
Note that the energy equation differs from that in \citet{Tricco2016}
because we implement divergence cleaninng on a Eulerian grid
and do not advect $\psi$ with the flow unlike in their smoothed
particle hydrodynamics approach. If translated to the
Eulerian form of the MHD equations, the approach of 
\citet{Tricco2016} would give rise to an extra advection
term in the evolution equation for $\psi$, which then
requires an additional term in the energy equation to ensure
energy conservation. In our Eulerian approach, these extra terms
are not needed.

Viscosity and resistivity are not included explicitly
in the ideal MHD approximation,
and only enter through
the discretisation of the MHD equations,
specifically
the spatial reconstruction, the computation of the Riemann fluxes, and the update of the
conserved quantities in the
case of reconstruct-solve-average
schemes.
\footnote{
For an attempt to quantify the numerical diffusivities,
see, e.g., \citet{rembiasz_17}. One should emphasise,
however, that that the concepts of numerical
and physical viscosity/resistivity must
be carefully distinguished. For one thing,
the numerical viscosity and resistivity
are problem-dependent as stressed by
\citet{rembiasz_17}. In the case of higher-order
reconstruction schemes, they are also
scale-dependent (and hence more akin
to higher-order hyperdiffusivities),
and they are affected by discrete
switches in the reconstruction scheme
and Riemann solver. Because of these
subtleties, the determination of the effective
(magnetic) Reynolds number of a simulation
is non-trivial
\citep[see Section~2.1.2 of][]{mueller_20}.
}
While this ``implicit large-eddy simulation'' (ILES) approach is
well-established for hydrodynamic turbulence \citep{grinstein_07}, the magnetohydrodynamic case
is more complicated because the behaviour in the relevant astrophysical regime of low (kinematic) viscosity
$\nu$ and
resistivity $\eta$ may still depend on the magnetic
Prandtl number $\mathrm{Pm}=\nu/\eta$. Different
from the regime later encountered in the supernova
core where $\mathrm{Pm}\gg 1$, oxygen shell burning is characterised by magnetic Prandtl numbers slightly below unity ($\mathrm{Pm} \sim 0.2$). With an effective Prandtl number of $\mathrm{Pm} \sim 1$
in the ILES approach \citep{federrath_11},
there is a potential concern that spurious small-scale dynamo amplification might arise due to the overestimation of the magnetic Prandtl number, or that saturation field strengths might be too high. The debate about the magnetic Prandtl number dependence in MHD turbulence is ongoing with insights from theory and simulations \citep[e.g.,][]{schekochihin_04,schekochihin_07,isakov_07,brandenburg_11,pietarila_10,sahoo_11,thaler_15,seshasayanan_17}, astrophysical observations \citep[e.g.,][]{christensen_09}, and laboratory experiments \citep{petrelis_07,monchaux_07}. There is, at the very least, the possibility of robust small-scale dynamo action and saturation governed
by balance between the inertial terms and Lorentz force terms down to the low values of $\mathrm{Pm}\sim 5\times 10^{-6}$ in liquid sodium experiments \citep{petrelis_07,monchaux_07}, provided that
both the hydrodynamic and magnetic Reynolds number are sufficiently high. An ILES approach that places the simulation into a ``universal'', strongly magnetised regime \citep{beresnyak_19} thus appears plausible. Moreover, even if the ILES approach were only to provide upper limits for magnetic fields and their effects on the flow, meaningful conclusions can still be drawn in the context of shell convection simulations.

The nuclear source terms are calculated with the 19-species nuclear reaction network of \citet{Weaver1978}. Neutrino cooling is ignored, since it becomes subdominant in the late pre-collapse phase as the contraction of the
shells speeds up nuclear burning.

The simulations are conducted on a grid
with $400\times128\times256$ zones in radius $r$, colatitude $\theta$, and longitude $\varphi$
with an exponential grid in $r$ and uniform
spacing in $\theta$ and $\varphi$.
To reduce computational costs, we excise the non-convective inner core up to $3000\,\mathrm{km}$ and replace the excised core with a point mass. The grid extends to a radius
of $50,000\, \mathrm{km}$ and includes a small part
of the silicon shell, the entire convective oxygen and neon shells, the non-convective carbon shell,
and parts of the helium shell. 
Simulations that include the entire iron core
and silicon shell are of course desirable in future,
but the treatment of nuclear quasi-equilibrium
in these regions is delicate and motivates simulations
in a shellular domain instead, with a sufficient ``buffer''
region between the convectively active shells and the inner boundary of the grid. Due to the significant buoyancy jump between
the silicon shell and the oxygen shell, the flow in the convective
shell is largely decoupled from that in the 
non-convective buffer region. Hydrodynamic
waves excited at the convective boundary and Alfv\'en waves along
field lines threading both regions can introduce some coupling, but these are excited by the active convective shell and not in the convectively quiet buffer region. No physical flux of waves 
\emph{into} the oxygen region from below is expected. Moreover,
there is no strong excitation of waves at the convective
boundary in the first place because of the low convective Mach number
of the flow. It turns out that coupling by Alfv\'en waves can be neglected even more safely than coupling by internal gravity waves because of the Alfv\'en number of the flow; as we shall see
the Alfv\'en number always stays well below unity.

In order to investigate the impact of magnetic fields on late-stage oxygen shell convection, we run a purely hydrodynamic, non-magnetic simulation and an MHD simulation. In the MHD simulation, we impose a homogeneous magnetic field with $B_z = 10^{8}\,\mathrm{G}$
parallel to the grid axis as initial conditions. We implement reflecting and periodic boundary conditions in $\theta$ and $\varphi$, respectively. For the hydrodynamic variables we use hydrostatic
extrapolation \citep{mueller_20} at the inner and outer boundary, and
impose strictly vanishing advective fluxes at the inner
boundary. 
Different from
\citet{mueller_16c}, we do \emph{not} contract
the inner boundary to follow the contraction and
collapse of the core.
This means that our models will not (and are not
intended to) provide an accurate representation of the
pre-collapse state of the particular $18 M_\odot$ star
that we are simulating. We would expect, e.g., that
for the particular $18 M_\odot$ model, the burning rate
and hence the convective velocities would increase
until the onset of collapse \citep{mueller_16c} due
to the contraction of the convective oxygen shell. As a consequence
of accelerating convection and flux compression, the magnetic
fields will likely also be somewhat higher at the onset of collapse.
The model is rather meant to reveal the physical principles
governing late-stage magnetoconvection in massive stars,
and to be \emph{representative} of the
typical conditions in oxygen burning shells with the understanding
that there are significant variations in convective Mach
number and shell geometry at the onset of collapse \citep{Collins2018},
which will also be reflected in the magnetic field strengths in the
interiors of supernova progenitors.

The inner and outer boundary conditions for the
magnetic fields are less trivial.
In simulations of magnetoconvection in the Sun, 
various choices such as vertical boundary conditions
($B_x=B_y=0$), radial boundary conditions ($B_\theta=B_\varphi=0$), vanishing tangential
electric fields or currents, perfect-conductor
boundary conditions, or extrapolation
to a potential solution have been employed
\citep[e.g.,][]{Thelen2000, Rempel2014,Kapyla2018}. 
Since our domain boundaries are separated from
the convective regions by shell interfaces with
significant buoyancy jumps, we opt for the
simplest choice of boundary conditions and merely
fix the magnetic fields in the ghost zones
to their initial values for a homogeneous vertical magnetic field. We argue that due to the buffer regions at our radial boundaries, and the lack of rotational shear, our choice of magnetic boundary conditions should not have a significant impact on the dynamically relevant regions of the star.

\section{Results}
\label{sec:results}

\subsection{Evolution of the Magnetic Field}
\label{subsec:magGeometry}

\begin{figure}
 \includegraphics[width=\columnwidth]{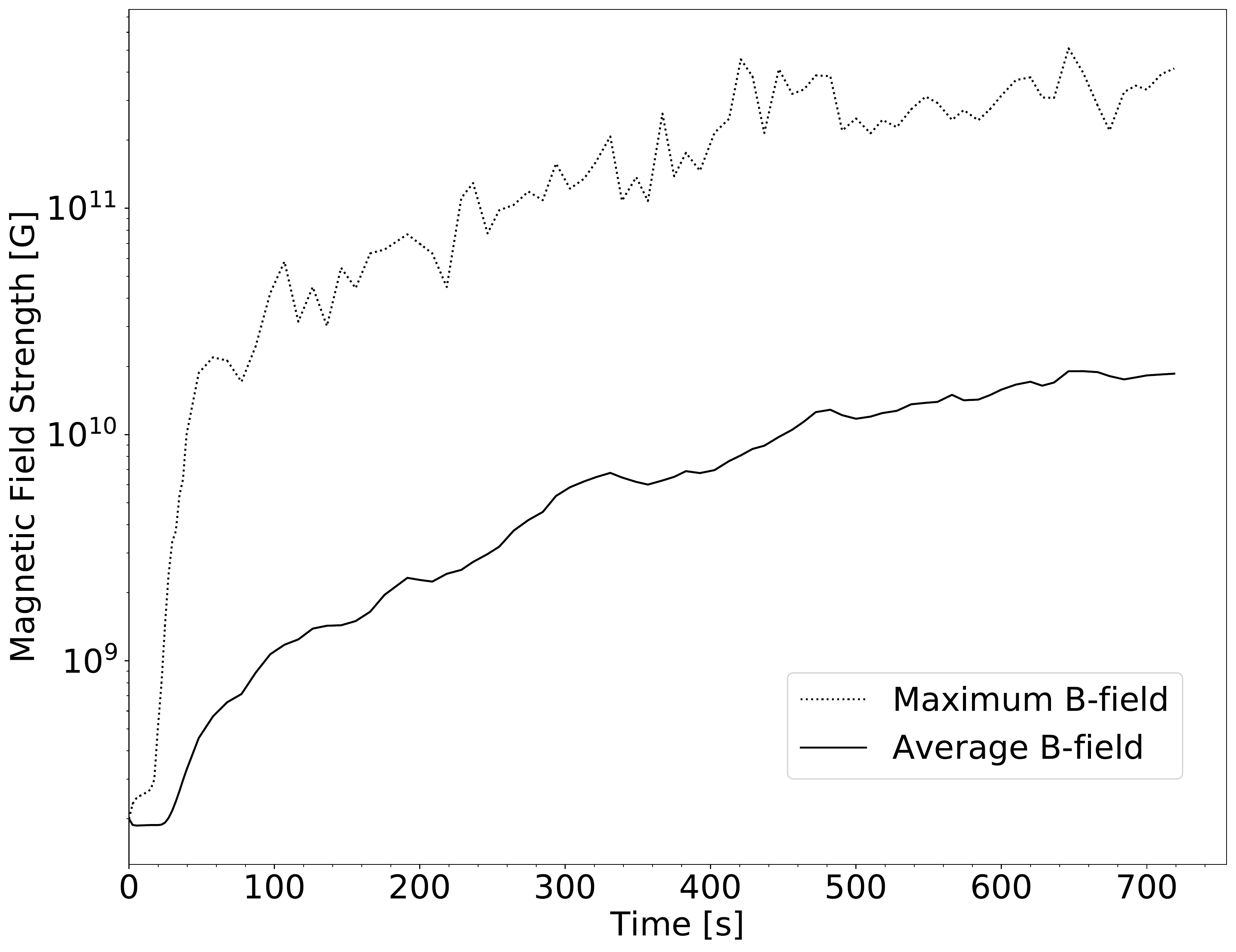}
 \caption{Evolution of the volume-averaged (solid) and maximum (dashed) magnetic field strength within the oxygen shell.
 }
 \label{fig:B_field}
\end{figure}

\begin{figure*}
 \includegraphics[width=\linewidth]{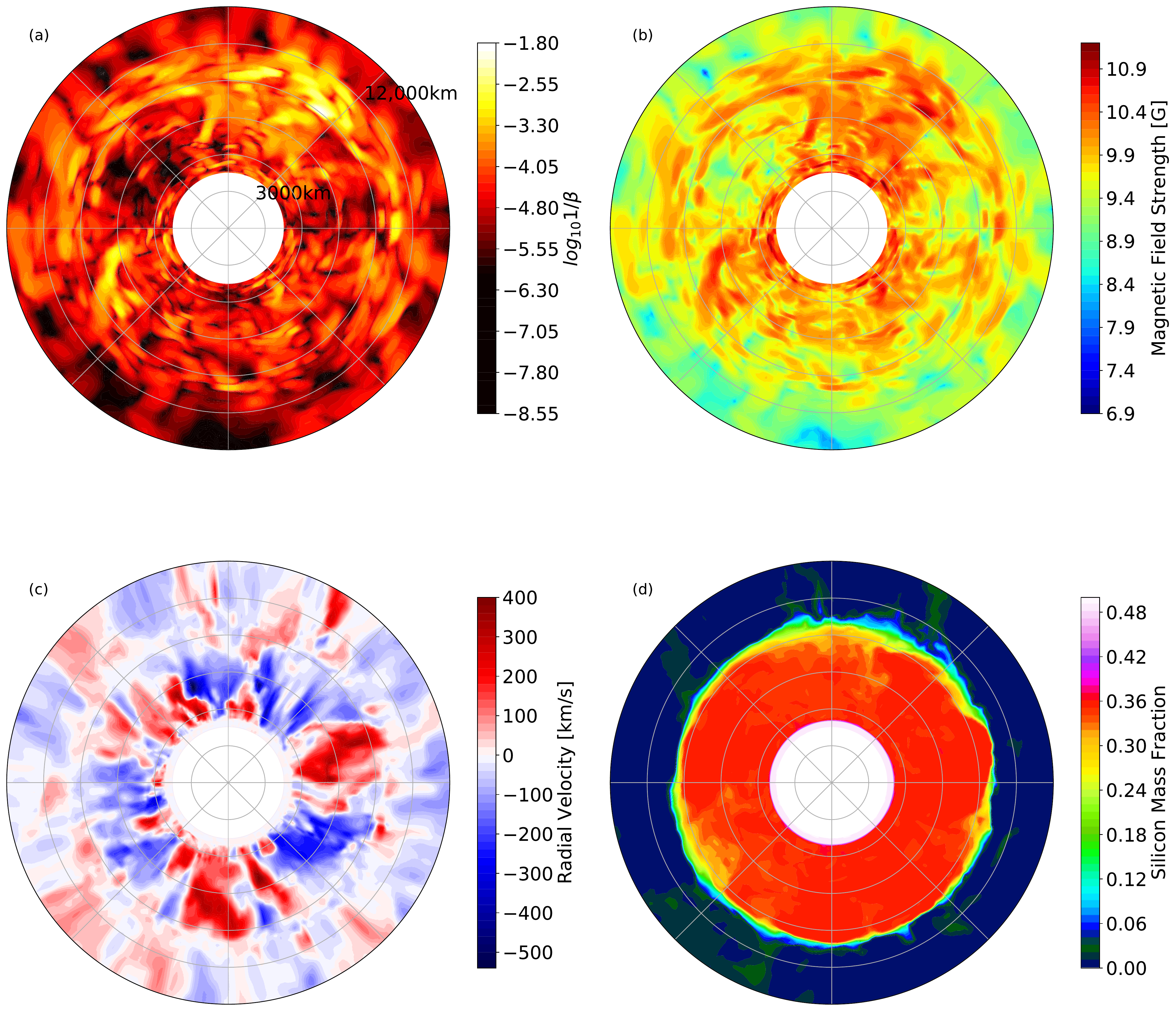}
 \caption{Snapshots of the equatorial plane in
 the MHD simulation at a time of  $500\, \mathrm{s}$,
 showing the inner part of the domain from the inner boundary at a radius of $3000 \, \mathrm{km}$ 
 out to $12,000\, \mathrm{km}$. The plotted part of the domain corresponds to the range $\mathrm{1.67\, M_\odot - 2.54\, M_\odot}$ in mass coordinate.} The panels display
 a) the ratio of magnetic to thermal pressure (i.e., inverse plasma-$\beta$), b) the magnitude of the magnetic field strength, c) the radial velocity and d) the silicon mass fraction.
 \label{fig:2Dslices}
\end{figure*}

Both the magnetised and non-magnetised model
were run for over 12 minutes of physical time, which
corresponds to about 17 convective turnover times.
Very soon after convection develops in the oxygen shell, the turbulent convective flow starts to rapidly amplify the magnetic fields in this region. To illustrate the growth
of the magnetic field we show the 
root-mean-square (RMS) average and maximum value of the 
magnetic field in the oxygen shell in Figure~\ref{fig:B_field}.The magnetic field, which we initialised at $10^8\, \mathrm{G}$, is amplified by over two orders of magnitude to over $10^{10} \,\mathrm{G}$ on average within the shell due to convective and turbulent motions. The average magnetic field strength in the shell is still increasing at the end of the simulation, but 
the growth rate has slowed down, likely indicating that the model is approaching some level of magnetic field saturation. While we cannot with certainty
extrapolate the growth dynamics without
simulating longer, it appears likely
that RMS saturation field strength will settle somewhere around $\mathord{\approx} 2 \times 10^{10}\,\mathrm{G}$. 

A closer look reveals that the magnetic field
is not amplified homogeneously throughout the
convective region. The convective eddies push the magnetic field lines against the  convective boundaries, where the fields are then more strongly amplified by shear flows. This is visualised in Figures~\ref{fig:2Dslices}a and \ref{fig:2Dslices}b, where both magnetic pressure and magnetic field strengths appear concentrated at the convective boundary. The maximum magnetic field strength 
shown in Figure~\ref{fig:B_field} (dashed line) therefore essentially mirrors the field at the inner boundary of the oxygen shell
(not to be confused with the inner grid boundary). We observe a very quick rise in magnetic field strength at the boundary at the beginning of the simulation between $20 \texttt{-} 50\,\mathrm{s}$ once
the convective flow is fully developed. The rate of growth of the maximum magnetic field strength after this increases at approximately the same rate as the average magnetic field strength.

\begin{figure}
    \includegraphics[width=\columnwidth]{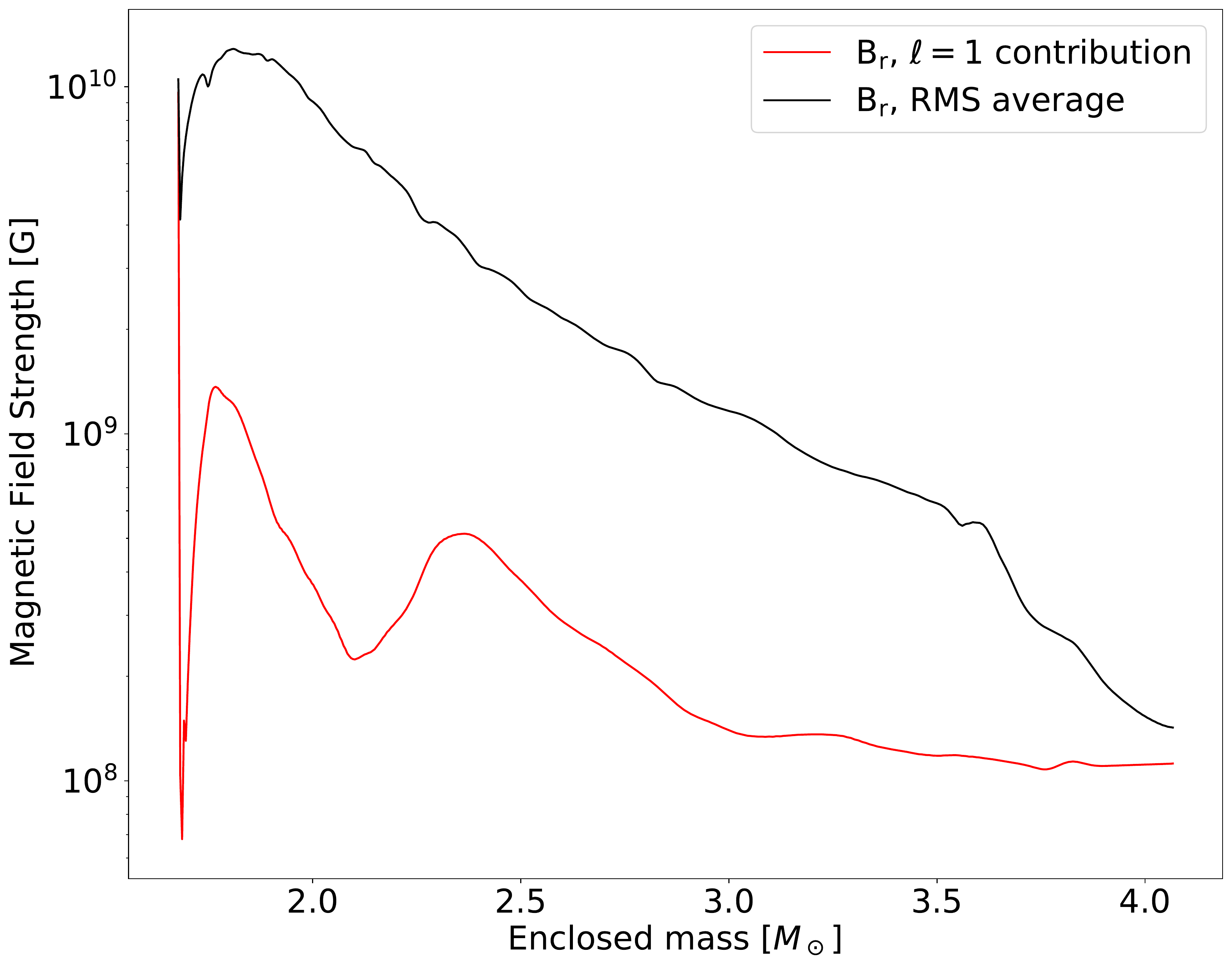}
    \caption{The angle-averaged root-mean-square (RMS) value
(black) and the dipole component (red) of the
radial magnetic field component $B_r$ 
as a function of mass coordinate at a time of $725\, \mathrm{s}$.
Note that the magnetic field strength drops dramatically
inside the thin, convectively stable  buffer region below
the oxygen shell, but exhibits a peak in the first radial zone,
which is merely a numerical artefact due to imperfect
boundary conditions.
    }
    \label{fig:Dipole}
\end{figure}

\begin{figure}
 \includegraphics[width=\linewidth]{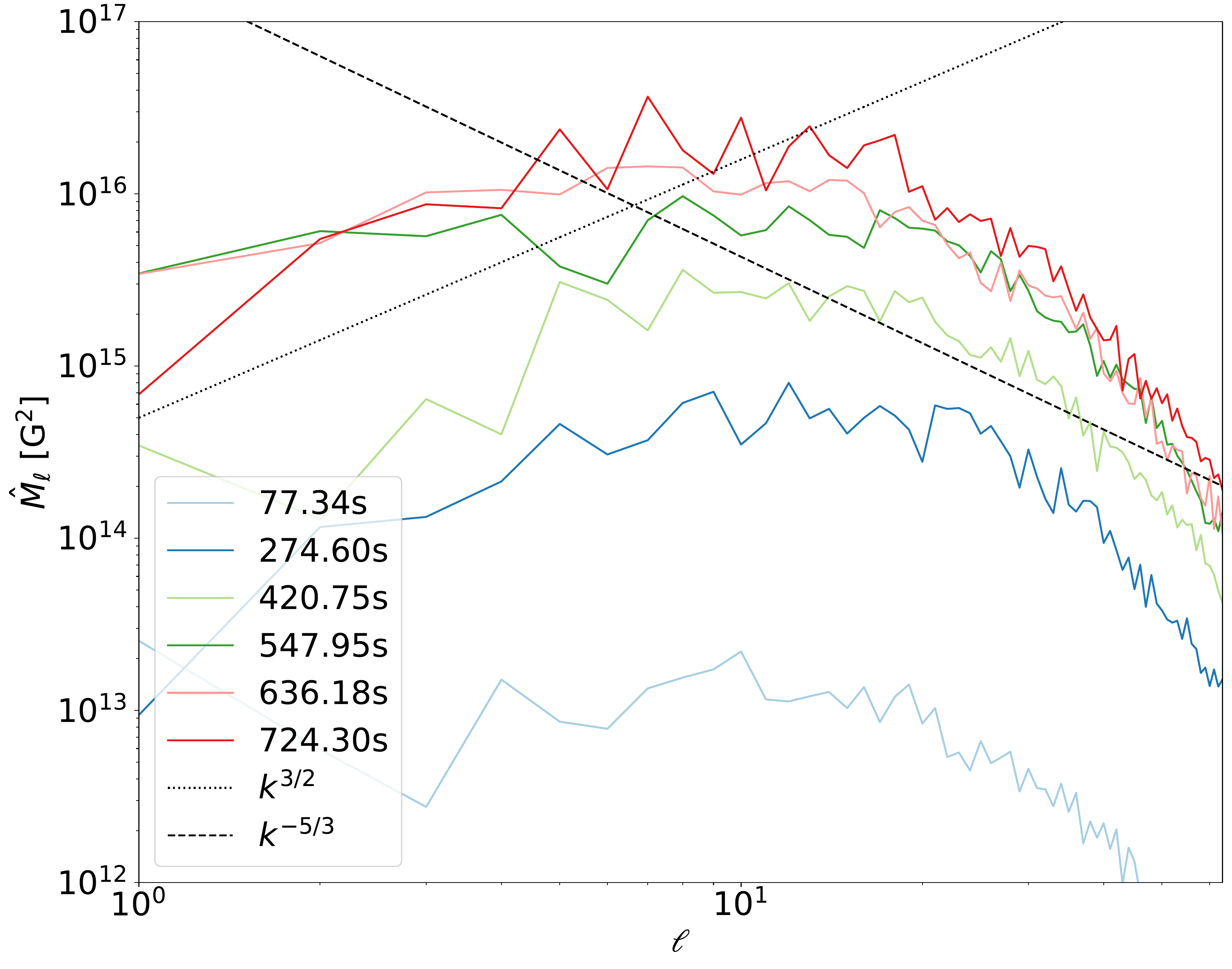}
 \caption{Power $\hat{M}_{\ell}^2$ in different multipoles $\ell$ of the radial field component
 of the magnetic field in the oxygen shell at different times. Dotted lines show the slopes
 of Kolmogorov ($k^{-5/3}$) and Kazantsev
 ($k^{3/2}$)  spectra. The low-wavenumber part of the spectrum
 is always distinctly flatter than a Kazantsev spectrum; at
 intermediate $\ell$, we see a Kolmogorov-like spectrum with a break
 around $\ell=30$ to a steeper slope
 in the dissipation range.
 }
 \label{fig:SphericalHarmonics}
\end{figure}

To characterise the geometric structure of the magnetic field, we show a radial profile of the dipole 
of the radial magnetic field component\footnote{Strictly speaking, the most rigorous way to extract the dipole component of the magnetic field
 would use a poloidal-toroidal decomposition 
$\mathbf{B}=
\nabla\times[\nabla \times (\mathcal{P}\mathbf{r})]
+
\nabla \times (\mathcal{T} \mathbf{r})$
where the scalar functions $\mathcal{P}$ and $\mathcal{T}$
describe the poloidal and toroidal parts of the field,
and consider \emph{all} components
$B_r$, $B_\theta$, and $B_\varphi$
arising from the $\ell=1$ component of
the poloidal scalar $\mathcal{P}$. Since the poloidal-toroidal decomposition cannot be reduced to a straightforward projection onto vector spherical harmonics,  this analysis is left to future papers.
} $B_r$ at the end of the simulation at $\approx 725\,\mathrm{s}$ (Figure~\ref{fig:Dipole}).
The dipole magnetic field in the convective regions is approximately an order of magnitude smaller than the RMS average radial magnetic field, aside from the first radial zone of our grid, where the dipole component is comparable to the total radial magnetic field. This behaviour at the grid boundary is likely an artefact of our choice of homogeneous magnetic fields at the inner boundary.  In general, however, the magnetic fields in the convective zones appear dominated by higher-order multipoles. Disregarding the dipole fields at the inner boundary, the dipole magnetic field of $\mathord{\approx} 10^9 \mathrm{G}$ or below lies in the upper range of observed dipole magnetic fields of white dwarfs \citep{Ferrario2015}, which have often been taken as best estimates for the dipole fields in the cores of massive stars.

\begin{figure}
 \includegraphics[width=\columnwidth]{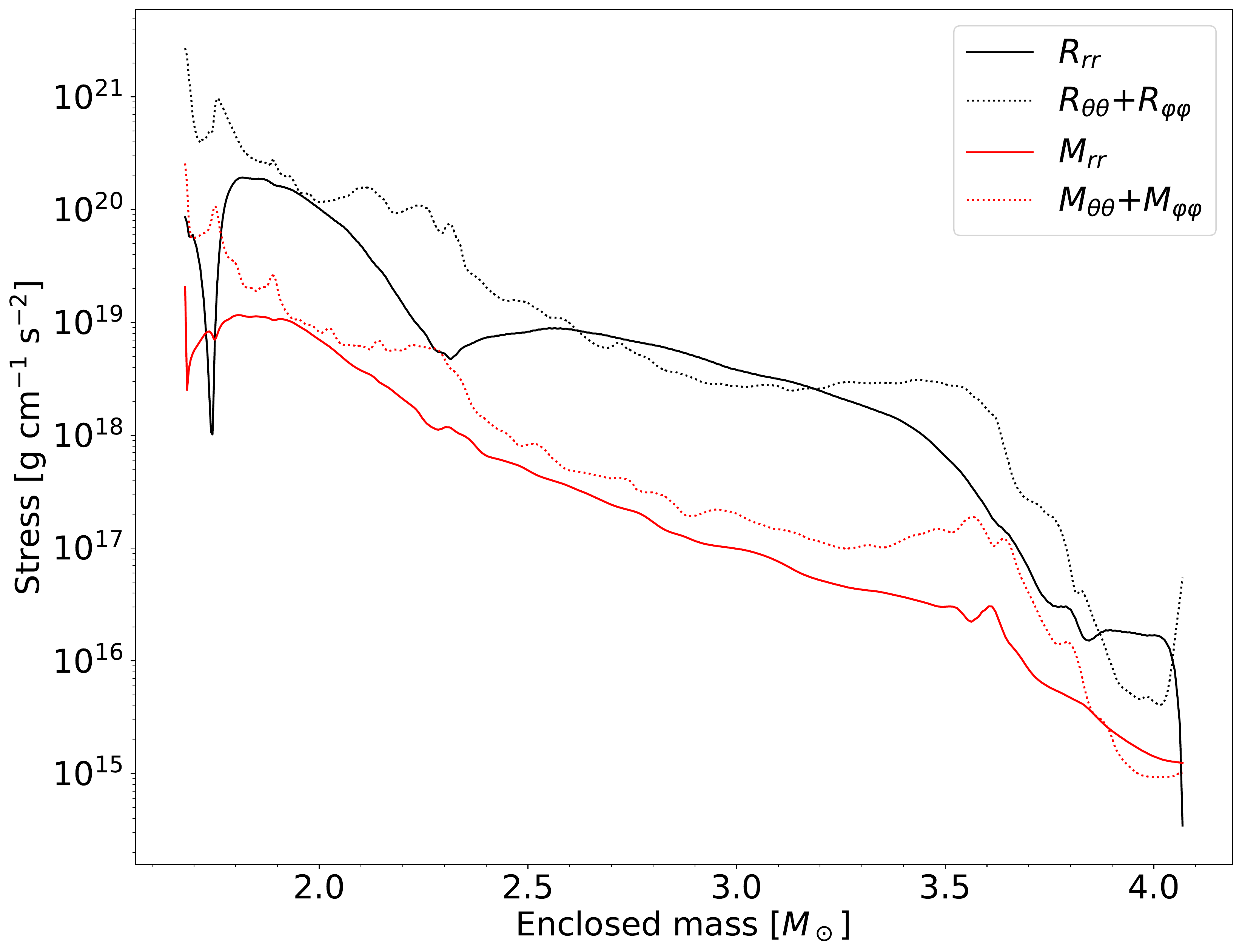}
 \caption{The radial  (solid) and non-radial (dashed)
 diagonal components of the 
 Reynolds stress tensor
 $R_{ij}$ (black) and  Maxwell 
 tensor $M_{ij}$ (red) 
 for the MHD convection model at $725\,\mathrm{s}$ as a function of enclosed mass.}
 \label{fig:Stress}
\end{figure}

To further illustrate the small-scale nature of
the magnetic field within the oxygen shell, we show angular power spectra, $\hat{M}_{\ell}$ of the radial field strength at different times as a function of the spherical harmonics degree $\ell$ inside the oxygen shell at a radius of $\mathord{\approx} 5000\mathrm{km}$ (Figure~\ref{fig:SphericalHarmonics}).
$\hat{M}_{\ell}$ is 
computed as:
\begin{eqnarray}
\hat{M}_\ell &=&
\frac{1}{8\pi}
\sum_{m=-\ell}^\ell \left|\int Y_{\ell m}^*(\theta,\varphi) B_r 
\mathrm{d} \Omega\right|^2.
\end{eqnarray}

Very early in the simulation, the spectrum shows a significant 
$\ell = 1$ dipole contribution, caused by our choice of homogeneous magnetic fields as an initial condition. It takes several convective turnovers before this contribution is no longer dominant. Throughout the evolution, the spectrum shows a broad plateau at small $\ell$ and a Kolmogorov-like slope at intermediate $\ell$,
which can be more clearly distinguished at late times when the first
break in the spectrum has moved towards lower $\ell$.
High-resolution studies are desirable to extend the inertial
range and confirm the development of a Kolmogorov spectrum at intermediate
wave numbers. The break in the spectrum moves towards smaller wave numbers and
the peak of the spectrum  shifts to larger scales from $\ell \approx  12$ to $ \ell \approx 7$. 
Simulations of field amplification by a  small-scale dynamo in isotropic turbulence often exhibit a Kazantsev spectrum
with power-law index $k^{-3/2}$ on large scales.
Our spectra show a distinctly flatter slope below the spectral speak, indicating that field amplification is subtly different from the standard picture of turbulent dynamo amplification.

This is borne out by a closer look at the magnetic
field distribution within the convective region. 
Somewhat similar to our recent simulation of field amplification by neutrino-driven convection in core-collapse supernovae \citep{Mueller2020}, field amplification does not happen homogeneously throughout the convective region and appears to be predominantly driven
by shear flows at the convective boundaries.
To illustrate this, we compare the 
spherically-averaged diagonal components of the
kinetic (Reynolds) and magnetic (Maxwell) stress tensors
$R_{ij}$ and $M_{ij}$
in the MHD model at the final time-step of the simulation at $\mathord{\approx} 725\,\mathrm{s}$ (Figure~\ref{fig:Stress}). 
$R_{ij}$ and $M_{ij}$ are computed as
\begin{eqnarray}
R_{ij}&=& \langle \rho v_i v_j\rangle, \\
M_{ij}&=& \frac{1}{8\pi}\langle B_i B_j\rangle, 
\end{eqnarray}
where angled brackets denote volume-weighted
averages.\footnote{Note that no explicit decomposition of the velocity field into fluctuating components and a spherically averaged  background state is required since the background state is hydrostatic.} 
The magnetic fields clearly remain well below
equipartition with the total turbulent kinetic
energy, and appear to converge to saturation
levels about one order of magnitude below, although
longer simulations will be required to confirm this.
The non-radial
diagonal components 
$R_{\theta\theta}+R_{\varphi\varphi}$ and
$M_{\theta\theta}+M_{\varphi\varphi}$
are generally higher than the respective radial components $R_{rr}$ and $M_{rr}$. Throughout
most of the domain, the Maxwell stresses are
considerably smaller than the Reynolds stresses,
but it is noteworthy that the profile of the non-radial components of $M_{ij}$ runs largely parallel to those of
$R_{ij}$, just with a difference of slightly more than
an order of magnitude. Peaks of the magnetic
stresses at the shell interfaces suggest
that field amplification is driven  by shear
flow at the convective boundaries. Convective motions
then transport the magnetic field into the interior
of the convective regions and also generate radial
field components. The humps of $M_{rr}$ within the convection zones are less pronounced than in $R_{rr}$, which indicates that little
amplification by convective updrafts
and downdrafts takes place within the convection region.

At the outer boundaries of
the oxygen and neon shell, we observe
rough equipartition  between
$R_{rr}$ and $M_{\theta\theta}+M_{\varphi\varphi}$. The fact that the growth of the field slows down once the model approaches
$M_{\theta\theta}+M_{\varphi\varphi}\approx R_{rr} $ suggests that this ``partial equipartition'' may determine the saturation field strength, but
the very different behaviour at
the inner boundary with 
 $M_{\theta\theta}+M_{\varphi\varphi}\gg R_{rr}$ argues against this. It is plausible, though, that the saturation
 field strength is (or rather will be) determined at the boundary. Linear stability analysis of the magnetised Kelvin-Helmholtz instability \citep[e.g.][]{chandrasekhar_61,Sen1963,Fejer1964,Miura1982,frank_96, keppens_99, ryu_00,obergaulinger_10, Liu2018} shows that shear instability, which is critical for efficiently generating small-scale fields, are suppressed by magnetic fields parallel to the direction of the shear flow as long as the Alfv\'en number
 is smaller than two. With kinetic fields well below equipartition
 (cp.~Figure~\ref{fig:Stress}), our models fall into the regime of low Alfv\'en
 number; hence the generation of non-radial fields at the boundaries may be self-limiting as a result of the stabilising impact
 of magnetic fields on the Kelvin-Helmholtz instability. A naive
 application of the principle of marginal stability would suggest saturation occurs
 once the Alfv\'en velocity at the boundary equals the shear velocity jump across the shell interface, which would imply
 $M_{\theta\theta}+M_{\varphi\varphi}\approx R_{\theta\theta}+R_{\varphi\varphi}$. Our simulation suggests that saturation probably occurs at significantly smaller values, and a more quantitative analysis of the saturation mechanism will clearly be required in future to elucidate the relation between the shear velocity (and possibly the width of the shear layer) and the saturation field strength. 
 
 \begin{figure}
 \includegraphics[width=\columnwidth]{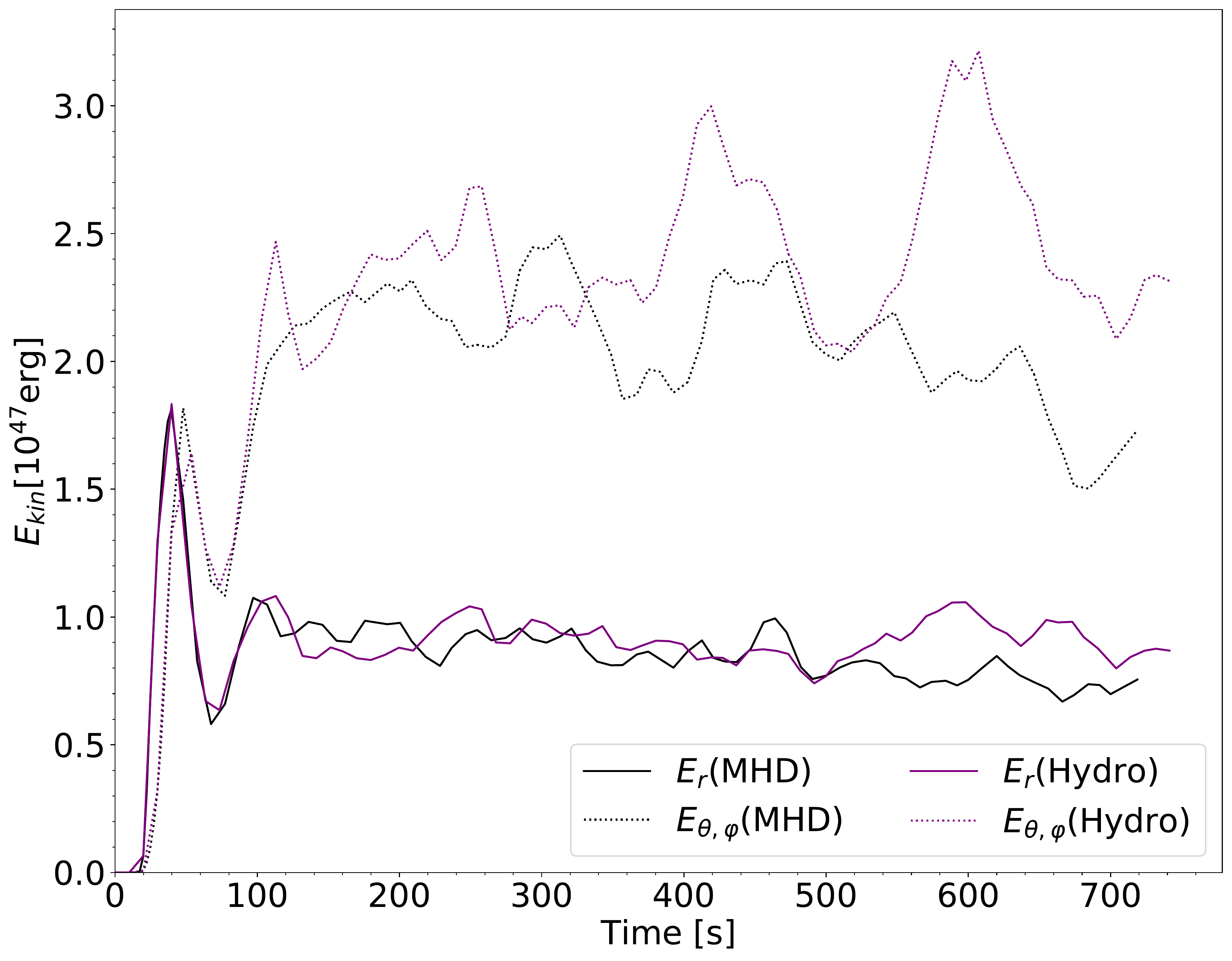}
 \caption{Evolution of the total radial (solid) and non-radial (dashed) convective kinetic energy within the oxygen shell for the purely hydrodynamic (purple) and MHD (black) model, respectively.}
 \label{fig:E_kin}
\end{figure}

\begin{figure*}
 \includegraphics[width=\linewidth]{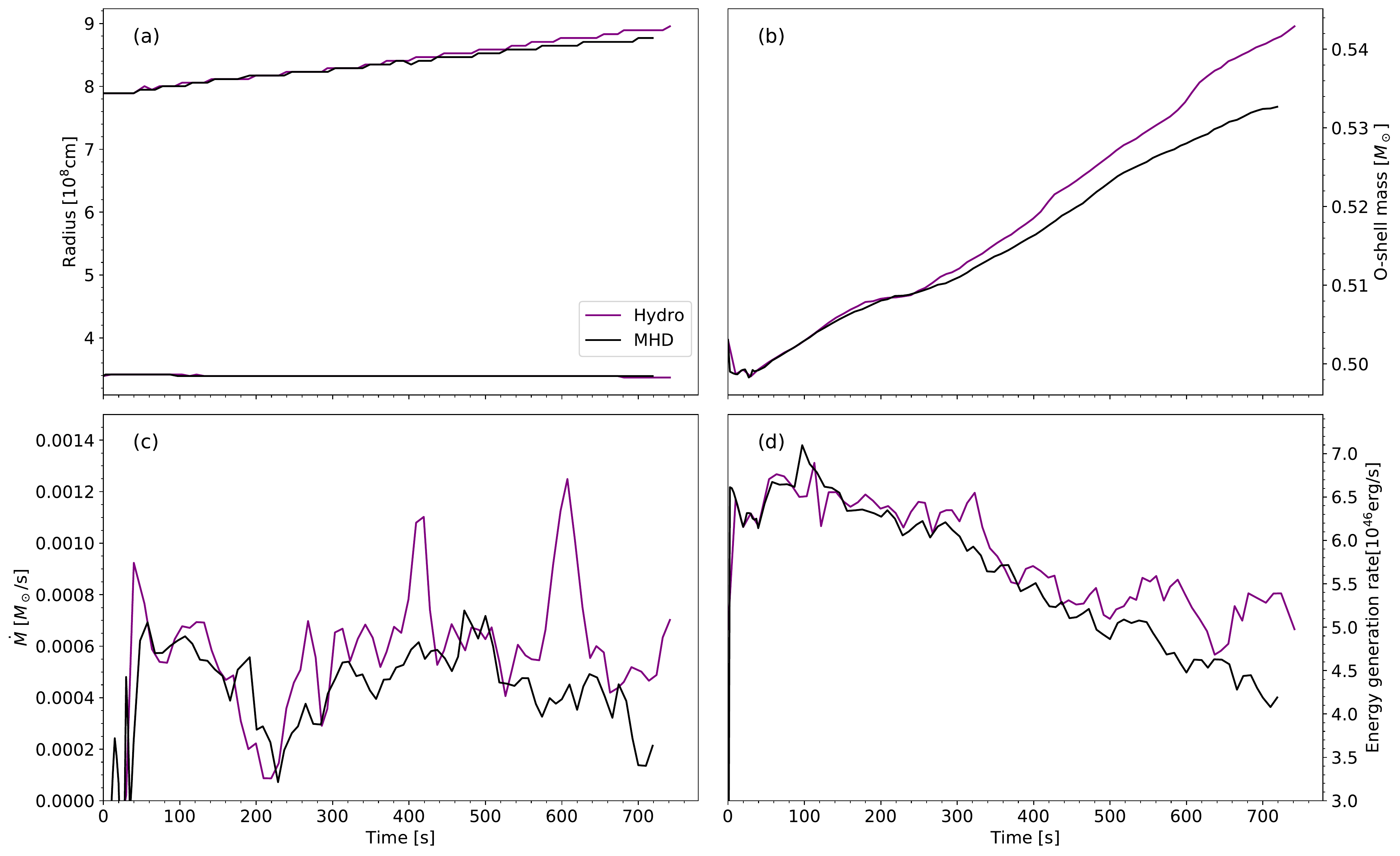}
 \caption{Effects of entrainment on the oxygen shell for both the MHD model (black) and the equivalent purely hydrodynamic model (purple).  The panels show
 a) the inner and outer  oxygen shell radii, b) the total mass within the oxygen shell, c) the entrainment rate $\dot{M}$ into the oxygen shell, and
 d) the volume-integrated nuclear energy generation rate throughout the shell.. 
}
 \label{fig:OShellValues}
\end{figure*}

\subsection{Impact of Magnetic Fields on Convective Boundary Mixing}
\label{subsec:entrainment}

The slowing growth of the magnetic field indicates that feedback effects on the flow should become important during the later phase of the simulation. It is particularly interesting to consider the effect of the strong fields tangential to the oxygen-neon shell interface on convective boundary mixing, though we also consider potential effects on the flow in the
interior of the convective regions.

To this end, we compare the MHD model to a purely hydrodynamic simulation of oxygen and neon shell convection. 
Figure~\ref{fig:E_kin} compares the total kinetic energy in convective motions in the
oxygen shell between the models.
The radial and non-radial components
$E_r$ and $E_{\theta, \varphi}$
of the kinetic energy are defined as
\begin{eqnarray}
E_r &=&\frac{1}{2} \int\limits_{r_- \leq r \leq r_+} \rho v_r^2 \,\mathrm{d}V,
\\
E_\mathrm{\theta, \varphi} &=&\frac{1}{2} \int\limits_{r_- \leq r \leq r_+} \rho (v_\mathrm{\theta}^2 + v_\mathrm{\varphi}^2 ) \,\mathrm{d}V,
\end{eqnarray}
where $r_-$ and $r_+$ are the inner and outer radii of the oxygen shell. We compute the inner
and outer shell radii $r_-$ and $r_+$ as the midpoints of the steep,
entropy slope between the oxygen shell and the silicon and neon shells below and above.  Due to shell
growth by entrainment, $r_-$ and $r_+$ 
are time-dependent (Figure~\ref{fig:OShellValues}a). 

For both models most of the turbulent kinetic energy is in the non-radial direction. This
is different from the rough equipartition 
$E_r \approx E_\mathrm{\theta, \varphi}$ seen in many simulations of buoyancy-driven convection
\citep{arnett_09}. There is, however,
no firm physical principle that dictates
such equipartition; indeed a shell burning simulation of the same $18\, M_\odot$ progenitor with the \textsc{Prometheus} code also showed significantly more kinetic energy in non-radial motions \citep{mueller_16c}. Ultimately, the high ratio $E_{\theta,\varphi}/E_r$ merely indicates that the fully developed flow happens to predominantly select eddies with larger extent in the horizontal than in the vertical direction.
\footnote{In the low-Mach number regime, the anelastic condition
$\nabla \cdot(\rho \mathbf{v})\approx 0$
implies a relation between the aspect
ratio of convective cells and the horizontal and vertical kinetic energy of any mode.}

Discounting stochastic variations, the radial component $E_r$ of the turbulent  kinetic energy for both the hydrodynamic and MHD model are  similar until the final $\mathord{\approx} 300\, \mathrm{s}$, at which point they start to deviate more significantly. Clearer differences 
appear in the non-radial component $E_{\theta,\varphi}$, with the hydro model showing irregular episodic bursts in kinetic energy which are not mirrored in the MHD model. Since the only difference in
the setup of the two runs is the presence or absence of magnetic fields, this points to feedback of the magnetic field on the flow as the underlying reason, unless the variations are purely stochastic. Closer examination
suggests that feedback of the magnetic field on the flow is
the more likely explanation, and
the nature of this feedback will become more apparent as we analyse convective boundary mixing in the models.

To this end, we first consider the
evolution of the boundaries $r_-$ and $r_+$ of the
oxygen shell and the total oxygen
shell mass $M_\mathrm{O}$. 
For computing the oxygen shell mass,
we take the (small) deviation of the boundaries from spherical symmetry into account more accurately than when
computing  $r_-$ and $r_+$, and integrate the mass contained in cells within
the entropy range  
$3.8 \texttt{-}5.2 \, k_\mathrm{B}/\mathrm{nucleon} $.
As shown by Figure~\ref{fig:OShellValues}
(panels a and b), the oxygen
shell in the non-magnetic model grows slightly, but perceptibly faster without magnetic fields than 
in the MHD model, starting
at a simulation time of about
$300\, \mathrm{s}$. The presence
of relatively strong magnetic fields
in the boundary layer apparently 
reduces entrainment in line
with the inhibiting effect of
 magnetic fields parallel to the
 flow on shear instabilities 
discussed in Section~\ref{subsec:magGeometry}.

The entrainment rate $\dot M=\ud M_\mathrm{O}/\ud t$
(Figure~\ref{fig:OShellValues}c)
is only slightly higher in the purely hydrodynamic model
most of the time, but the entrainment rate exhibits
occasional spikes, which do not occur in the MHD model. We note that these spikes in $\dot{M}$ occur at around the same times as the bursts in non-radial kinetic energy (Figure~\ref{fig:E_kin}). It appears that the
stabilisation of the boundary by magnetic fields
mostly suppresses rarer, but more
powerful entrainment events
that mix bigger lumps of material into
the oxygen shell. A comparison with the
shell burning simulations of \citet{mueller_16c} provides confidence that this effect is robust.
Because \citet{mueller_16c} contract the
inner boundary condition, convection grows more vigorous with time and their entrainment rates
are higher, adding about $0.05\, M_\odot$ to the
oxygen shell within $300\, \mathrm{s}$. 
Their resolution test (Figure~20 in \citealt{mueller_16c}) only showed variations of $0.004\, M_\odot$ in the entrained mass
between different runs of the same progenitor model. In our
simulations, the oxygen shell only grows by
about $0.03\, M_\odot$ from $300\, \mathrm{s}$
to $700 \, \mathrm{s}$ (i.e, when the magnetic
field does not grow substantially any more),
yet we find a difference in shell growth
of about $0.01\, M_\odot$ between the
magnetic and non-magnetic model. It therefore
seems likely that there is indeed an appreciable systematic effect of magnetic fields on entrainment. Our simulations qualitatively illustrate such an effect,
but models at higher resolution are of course desirable
to more reliably quantify the impact of magnetic fields on turbulent entrainment.
The resolution requirements for accurate entrainment
rates are hard to quantify. Findings from extant numerical studies of the magnetised Kelvin-Helmholtz instability \citep[e.g.,][]{keppens_99} cannot be easily
transferred because of important physical differences (e.g., the lack
of buoyant stabilisation of the boundary). In purely
hydrodynamic simulations of entrainment during oxygen shell burning, resolving the
shear layer between the oxygen and neon shell with $\mathord{\approx} 15\texttt{-}22$ radial zones
as in our model was found to be sufficient for quantitatively credible,
but not fully converged results \citep{mueller_16c}.

\begin{figure}
 \includegraphics[width=\columnwidth]{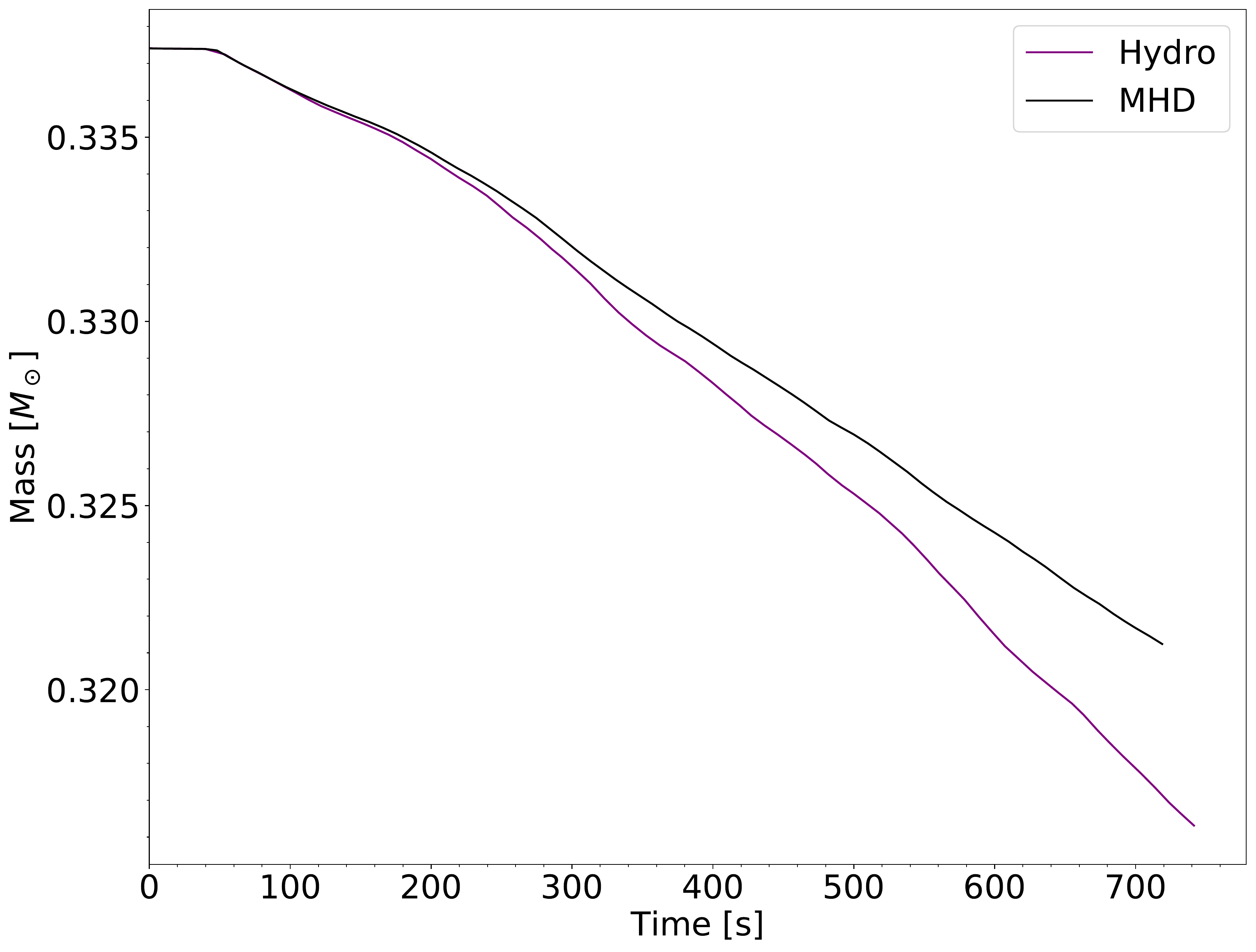}
 \caption{Total mass of neon in the entire computational domain for the purely hydrodynamic simulation (purple) and the MHD simulation (black).}
  \label{fig:NeonMass}
\end{figure}

The different entrainment rate also explains long-term, time-averaged differences in convective kinetic energy between the two simulations. The convective
kinetic energy is determined by the total nuclear
energy generation rate, which is shown in 
Figure~\ref{fig:OShellValues}d.
Both models show an overall trend downwards over time in nuclear energy generation, which can be ascribed
to a slight thermal expansion of the shell and the depletion of fuel. After about $200\,\mathrm{s}$, the purely
hydrodynamics model exhibits a higher energy
generation rate, which persists until the end of the simulations and becomes more pronounced.
The higher energy generation rate
in the non-magnetic model is indeed mirrored
by a stronger decrease in the neon mass on the grid (Figure~\ref{fig:NeonMass}) of about
the right amount to explain the average
difference in energy generation rate
at late times. We note, however, that
the strong episodic entrainment events
in the purely hydrodynamic model are 
\emph{not} associated with an immediate
increase in nuclear energy generation rate.
This is mainly because the energy release from the dissociation of the entrained neon is delayed and spread out in time as the entrained material is
diluted and eventually mixed down to regions
of sufficiently high temperature.

The overall effect of magnetic fields
on the bulk flow is rather modest.
We do not see a similarly strong quenching
of the convective flow by magnetic fields
as in  recent MHD simulations of the solar convection zone \citep{hotta_15}, who reported
a reduction of convective velocities
by up to $50\%$ at the base of the convection zone. Such an effect is not expected as long
as the magnetic fields stay well below kinetic
equipartition in the interior of convective shells as in our MHD simulation. Longer simulations will be required to confirm that  magnetic fields during late shell burning stages indeed remain ``weak'' by comparison and do not 
appreciably influence the bulk dynamics of convection. 

\section{Conclusions}
\label{sec:conclusion}
We investigated the amplification and saturation of magnetic fields during convective oxygen shell burning and the backreaction of the field on the convective flow by conducting a 3D MHD simulation
and a purely hydrodynamic simulation of an $18\, M_\odot$ progenitor shortly before core collapse.
The simulations were run for about 12 minutes of physical time (corresponding to about
17 convective turnover times), at which point field amplification
has slowed down considerably, though a quasi-stationary state has not yet been fully established.

The magnetic field in the oxygen shell is amplified to 
$\mathord{\sim}10^{10}\, \mathrm{G}$ and dominated
by small-to-medium-scale structures with angular
wavenumber $\ell \sim 7$. The dipole component
is considerably smaller with $\mathord{\sim}10^9\, \mathrm{G}$ near the inner boundary of
the oxygen shell and less further outside.
The profiles of the radial and non-radial diagonal components $M_{rr}$ and  $M_{\theta\theta}+M_{\varphi\varphi}$ of
the Maxwell stress tensor mirror the corresponding
components $R_{rr}$ and 
$R_{\theta\theta}+R_{\varphi\varphi}$ of the Reynolds stress tensor, but remain about an order of magnitude
smaller, i.e., kinetic equipartition is not reached.
However, $M_{\theta\theta}+M_{\varphi\varphi}$
can approach or exceed the radial
 component $R_{rr}$ at the convective boundaries.
The saturation mechanism for field amplification needs
to be studied in more detail, but we speculate that
saturation is mediated by the inhibiting effect of non-radial magnetic fields on shear instabilities at shell boundaries, which appear to be the primary driver of field amplification.

We find that magnetic fields do not have an appreciable effect on the interior flow inside the oxygen shell, but
observe a moderate reduction of turbulent entrainment at
the oxygen-neon shell boundary in the presence of magnetic fields. Magnetic fields appear to suppress stronger episodic entrainment events, although they do not quench entrainment completely. Through the reduced entrainment rate, magnetic fields also indirectly affect the dynamics inside the convective region slightly because they reduce the energy release through the dissociation of ingested neon, which results in slightly smaller convective velocities in the MHD model.

Our findings have important implications for core-collapse supernova modelling. We predict initial fields in the oxygen shell of non-rotating progenitors that are significantly stronger than assumed in our recent simulation of a neutrino-driven explosion aided by dynamo-generated magnetic fields \citep{Mueller2020}. With relatively strong seed-fields, there is likely less of a delay until magnetic fields can contribute the additional ``boost'' to neutrino heating  
and purely hydrodynamic instabilities seen in
\citet{Mueller2020}. This should further contribute to the robustness of the neutrino-driven mechanism for non-rotating and slowly-rotating massive stars. Our simulations also suggest that the perturbation-aided mechanism \citep{couch_13,mueller_15a} will not be substantially affected by the inclusion of magnetic fields. Since magnetic fields do not become strong enough to substantially alter the bulk flow inside the convective region, the
convective velocities and eddy scales as the key parameters for perturbation-aided explosions remain largely unchanged.

The implications of our results for neutron star magnetic fields are more difficult to evaluate since the observable fields will, to a large degree, be set by processes during and after the supernova and cannot be simply extrapolated from the progenitor stage by magnetic flux conservation. That said, dipole fields of order $10^9\, \mathrm{G}$ at the
base of the oxygen shell -- which is likely to end up as the neutron star surface region -- are not in overt conflict with dipole fields
of order $10^{13}\, \mathrm{G}$ in many young pulsars inside supernova remnants \citep{Enoto2019}. However, considering the
relatively strong small-scale fields
of $\mathord{\sim}10^{10}\, \mathrm{G}$
with peak values over $10^{11}\,\mathrm{G}$,
it may prove difficult to produce neutron stars without strong small-scale fields at the surface. There is a clear need for an integrated approach towards the evolution of magnetic fields from the progenitor phase through the supernova and into the compact remnant phase in order to fully grasp the implications of the current simulations.

Evidently, further follow-up studies are also needed on the final evolutionary phases of supernova progenitors.
Longer simulations and resolution studies will be required to better address issues like the saturation field strength, the saturation mechanism, and the impact of magnetic fields on turbulent entrainment.
Full models that include the core and self-consistently follow the evolution of the star to collapse will be required to generate
accurate 3D initial conditions for supernova simulations.
The critical issue of non-ideal effects and the behaviour of turbulent magnetoconvection for magnetic Prandtl numbers slightly smaller than one at very high Reynolds numbers deserves particular consideration. Some findings of our ``optimistic'' approach based on the ideal MHD approximation should, however, prove robust, such as the modest effect of magnetic fields on the convective bulk flow and hence the reliability of purely hydrodynamic models \citep{couch_15,mueller_16c,Yadav2020,mueller_20,Fields2020a} and even 1D mixing-length theory \citep{Collins2018} to  predict pre-collapse perturbations in supernova progenitors.
Future 3D simulations will also have to address rotation and its interplay with convection and magnetic fields.

\section*{Acknowledgements}
We thank A.~Heger for fruitful conversations.
BM was supported by ARC Future Fellowship FT160100035. 
We acknowledge computer time allocations
from NCMAS (project fh6) and ASTAC. This research was undertaken with the assistance of resources and services from the National Computational Infrastructure (NCI), which is supported by the
Australian Government.  It was supported by resources provided by the Pawsey Supercomputing Centre with funding from the Australian Government and the Government of Western Australia.

\section*{Data Availability}
The data underlying this article will be shared on reasonable request to the  authors, subject to considerations of intellectual property law.

\appendix
\section{Energy-conserving formulation of hyperbolic divergence cleaning}
\label{app:eglm}
The original extended generalised Lagrange multiplier formulation
of the MHD equations of \citet{Dedner2002} reads
(without non-adiabatic energy source/sink terms and the advection
equations for the mass fractions),
\begin{eqnarray}
\label{eq:glm1}
\partial_t \rho
+\nabla \cdot \rho \mathbf{v}
&=&
0,
\\
\partial_t (\rho \mathbf v)
+\nabla \cdot \left(\rho \mathbf{v}\mathbf{v}-
\frac{\mathbf{B} \mathbf{B}}{4\pi}
+P_\mathrm{t}\mathcal{I}
\right)
&=&
\rho \mathbf{g}
-
\frac{(\nabla \cdot\mathbf{B}) \mathbf{B}}{4\pi}
,
\\
\partial_t e+
\nabla \cdot 
\left[(e+P_\mathrm{t})\mathbf{u}
-\frac{\mathbf{B} (\mathbf{v}\cdot\mathbf{B})
}{4\pi}
\right]
&=&
\rho \mathbf{g}\cdot \mathbf{v}
-\frac{\mathbf{B} \cdot \nabla \psi}{4\pi}
,
\\
\partial_t \mathbf{B} +\nabla \cdot (\mathbf{v}\mathbf{B}-\mathbf{B}\mathbf{v})
+\nabla {\psi}
&=&0,
\\
\label{eq:glm_last}
\partial_t {\psi}
+c_\mathrm{h}^2 \nabla \cdot \mathbf{B}
&=&-{\psi}/\tau.
\end{eqnarray}
where $e=\rho \left(\epsilon+v^2/2\right)+B^2/(8\pi)$ is the standard
expression for the total internal, kinetic, and magnetic energy
of the fluid. This system includes  an extra source
term $-\mathbf{B} \cdot \nabla \psi/(4\pi)$ in the energy equation.
To ensure total energy conservation, one must also take into
account the energy carried by the cleaning field $\psi$,
which can be worked out as $e_\psi=\psi^2/(8\pi c_\mathrm{h}^2)$
\citep{tricco_12}.  
\citet{Tricco2016} noted that
that the original system of equations
(\ref{eq:glm1}--\ref{eq:glm_last}) of
\citet{Dedner2002} no longer guarantees total energy conservation if
a variable cleaning speed is used.
\footnote{
In addition, energy conservation is also violated
if one reformulates the equation for $\psi$
using a convective derivative, $\ud \psi/\ud t
+c_\mathrm{h}^2 \nabla \cdot \mathbf{B}
=
-{\psi}/\tau$, which can be rectified by including
an additional source term in the energy equation
\citep{tricco_12}. This problem is immaterial, however,
as long as $\psi$ is not advected with the fluid in a Eulerian
formulation.
}
They show that total energy conservation is recovered
if one introduces a rescaled cleaning field $\hat{\psi}=\psi/c_\mathrm{h}$
and solves a slightly modified system of equations,
\begin{eqnarray}
\partial_t \rho
+\nabla \cdot \rho \mathbf{v}
&=&
0,
\\
\partial_t (\rho \mathbf v)
+\nabla \cdot \left(\rho \mathbf{v}\mathbf{v}-
\frac{\mathbf{B} \mathbf{B}}{4\pi}
+P_\mathrm{t}\mathcal{I}
\right)
&=&
\rho \mathbf{g}
-
\frac{(\nabla \cdot\mathbf{B}) \mathbf{B}}{4\pi}
,
\\
\label{eq:energy}
\partial_t e+
\nabla \cdot 
\left[(e+P_\mathrm{t})\mathbf{u}
-\frac{\mathbf{B} (\mathbf{v}\cdot\mathbf{B})}{4\pi}
\right]
&=&
\rho \mathbf{g}\cdot \mathbf{v}
-\frac{\mathbf{B} \cdot \nabla (c_\mathrm{h}\hat{\psi})}{4\pi}
,
\\
\label{eq:b_field}
\partial_t \mathbf{B} +\nabla \cdot (\mathbf{v}\mathbf{B}-\mathbf{B}\mathbf{v})
+\nabla (c_\mathrm{h}\hat{\psi}),
&=&0,
\\
\label{eq:psi_field}
\partial_t \hat{\psi}
+c_\mathrm{h} \nabla \cdot \mathbf{B}
&=&-\hat{\psi}/\tau,
\end{eqnarray}
It is evident that rescaling the cleaning field does not alter the characteristic
structure of the system, but the more symmetric formulation of Equations~(\ref{eq:b_field})
and (\ref{eq:psi_field}) with identical dimensions for $\hat{\psi}$ and $\mathbf{B}$ allows us to explicitly cast the energy equation in conservation form.
The term $c_\mathrm{h} \nabla \cdot \mathbf{B}$
in  Equation~(\ref{eq:psi_field}) gives a contribution
(denoted by a subscript ``h'') to the time derivative of the energy associated with the cleaning field,
\begin{equation}
\label{eq:epsi}
\left(
\frac{\partial (\hat{\psi}^2/2)}{\partial t}
\right)_\mathrm{h}
=-c_\mathrm{h}
\hat{\psi}  \nabla \cdot \mathbf{B}.
\end{equation}
Because of the rescaling of the cleaning field, no time derivative of
the cleaning speed $c_\mathrm{h}$ appears, which is critical for 
writing the energy equation in a manifestly conservative form.
Adding Equations~(\ref{eq:energy}) and (\ref{eq:epsi}) yields
\begin{equation}
    \partial_t \hat{e}+
\nabla \cdot 
\left[(e+P_\mathrm{t})\mathbf{u}
-\frac{\mathbf{B} (\mathbf{v}\cdot\mathbf{B})}{4\pi}
\right]
=
\rho \mathbf{g}\cdot \mathbf{v}
-\frac{\mathbf{B} \cdot \nabla (c_\mathrm{h}\hat{\psi})+c_\mathrm{h} \hat{\psi} \nabla\cdot \mathbf{B}}{4\pi}
,
\end{equation}
where $\hat{e}=\rho (\epsilon+v^2/2)+(B^2+\hat{\psi}^2)/(8\pi)$.
The cleaning terms on the right-hand side can be written as a divergence and
hence absorbed in the energy flux,
\begin{equation}
\label{eq:etot}
\partial_t \hat{e}+
\nabla \cdot 
\left[(e+P_\mathrm{t})\mathbf{u}
-\frac{\mathbf{B} (\mathbf{v}\cdot\mathbf{B})
-c_\mathrm{h} \hat{\psi} \mathbf{B}}{4\pi}
\right]
=
\rho \mathbf{g}\cdot \mathbf{v}.
\end{equation}
As noted by \citet{Tricco2016}, the damping term for the cleaning field will effectively
add to dissipation into thermal energy in such an approach, but they found the
additional dissipation insignificant in practice compared to other
sources of numerical dissipation. 

\color{black}



\bibliographystyle{mnras}
\bibliography{paper}


\bsp	
\label{lastpage}
\end{document}